\documentclass[12pt]{article}

\usepackage[margin=1in]{geometry}
\usepackage{graphicx}
\usepackage{hyperref}
\hypersetup{
    colorlinks,
    linkcolor={blue},
    citecolor={blue},
    urlcolor={blue}
}

\usepackage{amsmath,amssymb,amsthm,amsfonts,mathrsfs}
\usepackage{xcolor}
\usepackage{enumitem}
\usepackage{booktabs,array,longtable,tabularx}
\usepackage[breakable,skins]{tcolorbox}
\usepackage[authoryear,round,compress]{natbib}
\usepackage[capitalise]{cleveref}
\usepackage{authblk}
\usepackage{xspace}
\usepackage{tikz}
\usetikzlibrary{arrows.meta,positioning,fit,calc,shapes.geometric}

\setlist[itemize]{leftmargin=2em}
\setlist[enumerate]{leftmargin=2em}

\newcommand{\E}{\mathbb{E}}
\newcommand{\Var}{\mathrm{Var}}

\newcommand{\titleString}{Reorganizing Graduate Statistical Inference for Modern Biostatistical Practice}

\newtcolorbox{pedagogicalgoal}{
    colback=gray!5,
    colframe=black!40,
    boxrule=0.5pt,
    arc=2pt,
    left=6pt,
    right=6pt,
    top=5pt,
    bottom=5pt
}

\begin{document}

\title{\Large \titleString}
\date{}

\author[1,2]{Raphael C. Kim}
\author[2]{Michele Santacatterina}

\affil[1]{Cornell Tech, Cornell University, New York, NY}
\affil[2]{Division of Biostatistics, Department of Population Health, New York University School of Medicine, New York, NY}

\maketitle

\begin{abstract}
First-year graduate courses in statistical inference within PhD programs in biostatistics often follow a classical mathematical statistics progression through probability models, data reduction, point estimation, testing and interval estimation, and asymptotic evaluation. This organization has clear pedagogical strengths, but may provide less guidance when the probability model is not the natural starting point. We propose a complementary target-first organization for a graduate statistical inference course in biostatistics. The course begins with the scientific question and target parameter, then introduces statistical model classes, estimation, first-order sampling behavior, inference, and efficiency within a common framework. Classical methods such as maximum likelihood, method of moments, estimating equations, confidence intervals, hypothesis tests, bootstrap and jackknife methods, and the Cramér--Rao bound are organized within a common inferential workflow. A central advantage of this organization is that the same workflow provides a natural bridge from classical inference to modern statistical problems, including flexible estimation using machine learning and artificial intelligence, policy learning, anytime-valid inference, and modern Bayesian methods.
\end{abstract}

\clearpage

\section{Introduction}

Many first-year graduate courses in statistical inference in biostatistics programs build on probability and follow a classical mathematical statistics progression: probability and models, data reduction, point estimation, testing and interval estimation, followed by asymptotic evaluation. This sequence has clear pedagogical advantages. Parametric families provide concrete probability models with tractable calculations, sufficiency and completeness formalize data reduction, likelihood and related methods provide systematic approaches to point estimation, and tests and confidence intervals introduce formal uncertainty quantification. This progression naturally encourages a \textit{model-first} view of inference: begin by specifying a parametric probability model for the observed data, then develop estimation and inference within that model. Texts such as \citet{CasellaBerger2002} follow this pedagogy closely. This sequence is intended to represent a common organizing pattern rather than to characterize every graduate inference course.

The model-first perspective is natural when a parametric model provides a useful starting point. Modern biostatistical problems, however, increasingly arise in settings where no such model is obvious. Electronic health records, genomics, imaging, and other complex data structures may require flexible, data-adaptive methods, including machine learning and artificial intelligence tools. The challenge is therefore not that classical inferential principles cease to apply, but that a model-first pedagogy provides less guidance about how to organize inference when the model is no longer the natural starting point. This may make it more difficult for students to connect classical theory with contemporary biostatistical practice.

A complementary way to organize statistical inference is to begin with the scientific question and define the statistical target, or quantity, that answers it. We refer to this as a \textit{target-first} view of inference. Once the target has been specified, students can consider what statistical model class or structure to impose on the data-generating distribution, how to construct an estimator of the target, and how to study its sampling behavior, uncertainty, and efficiency. This places classical methods within a common inferential workflow that can remain useful as estimation becomes more flexible and statistical problems evolve.

A target-first organization can also help students connect frequentist and Bayesian approaches. Both can begin from the same scientific question and statistical target, while differing in how inference about that target is constructed and uncertainty is quantified. Framing the two approaches around a common target can therefore make their similarities and differences easier to understand.

To better prepare graduate students for contemporary biostatistical practice, we propose reorganizing the first graduate course in statistical inference around a target-first workflow. The contribution is pedagogical rather than methodological: classical inferential ideas remain central, but are reorganized around a common sequence linking scientific questions, statistical targets, models, estimation, sampling behavior, and inference. A central advantage of this organization is that the same sequence provides a natural bridge from classical inference to modern problems involving flexible machine-learning and artificial-intelligence methods, policy and reinforcement learning, sequential inference, and modern Bayesian methods.

\section{A Target-First Framework for Statistical Inference}

At a high level, the course is organized around a sequence that begins with the scientific question and target parameter, then proceeds through model class choice, estimation, sampling behavior, inference, and efficiency.

Within this organization, classical topics such as  maximum likelihood estimation, estimating equations, hypothesis testing, confidence intervals, and the Cramér--Rao bound are introduced as components of a broader inferential workflow rather than as the organizing framework itself. The same structure also provides natural places for bootstrap and jackknife methods and can subsequently be reused to introduce multiple testing, regression and working models, and basic Bayesian inference.

\begin{figure}[h!]
\centering

\begin{minipage}[t]{0.42\linewidth}
\vspace{0pt}
\centering

\begin{tikzpicture}[
    node distance=0.55cm,
    box/.style={
        draw, rectangle, rounded corners,
        align=center,
        minimum width=3.8cm,
        minimum height=0.80cm,
        inner sep=2pt,
        font=\footnotesize
    },
    arrow/.style={->, thick}
]

\node[box] (prob) {
\textbf{Probability and Models} \\
{\scriptsize distributions, parametric families}
};

\node[box, below=of prob] (structure) {
\textbf{Data Reduction} \\
{\scriptsize sufficiency, completeness}
};

\node[box, below=of structure] (estimation) {
\textbf{Point Estimation} \\
{\scriptsize likelihood, MLE, method of moments}
};

\node[box, below=of estimation] (testing) {
\textbf{Testing and Interval Estimation} \\
{\scriptsize tests, confidence intervals}
};

\node[box, below=of testing] (asymp) {
\textbf{Asymptotic Evaluation} \\
{\scriptsize consistency, limiting distributions}
};

\draw[arrow] (prob) -- (structure);
\draw[arrow] (structure) -- (estimation);
\draw[arrow] (estimation) -- (testing);
\draw[arrow] (testing) -- (asymp);

\end{tikzpicture}

\vspace{0.3cm}
\footnotesize \textbf{Classical model-first sequence}

\end{minipage}
\hfill
\begin{minipage}[t]{0.56\linewidth}
\vspace{0pt}
\centering

\begin{tikzpicture}[
    node distance=0.42cm,
    box/.style={
        draw, rectangle, rounded corners,
        align=center,
        minimum width=4.1cm,
        minimum height=0.70cm,
        inner sep=2pt,
        font=\footnotesize
    },
    arrow/.style={->, thick}
]

\node[box] (question) {
\textbf{Scientific Question}
};

\node[box, below=of question] (target) {
\textbf{Target Parameter} \\
{\scriptsize $\psi_0=\Psi(P_0)$ \quad (\cref{sec:target})}
};

\node[box, below=of target] (model) {
\textbf{Statistical Model Class} \\
{\scriptsize $P_0\in\mathcal M$ \quad (\cref{sec:models})}
};

\node[box, below=of model] (sample) {
\textbf{Observed Sample} \\
{\scriptsize $O_1,\ldots,O_n\sim P_0$}
};

\node[box, below=of sample] (estimation) {
\textbf{Estimation} \\
{\scriptsize plug-in, Z-estimation, M-estimation \quad (\cref{sec:estimation})}
};

\node[box, below=of estimation] (firstorder) {
\textbf{First-Order Behavior and Inference} \\
{\scriptsize target sensitivity, estimator behavior,}\\
{\scriptsize regularity, influence functions, uncertainty \quad (\cref{sec:sampling})}
};

\node[box, below=of firstorder] (efficiency) {
\textbf{Efficiency} \\
{\scriptsize first-order variance, Cramér--Rao bound \quad (\cref{sec:efficiency})}
};

\draw[arrow] (question) -- (target);
\draw[arrow] (target) -- (model);
\draw[arrow] (model) -- (sample);
\draw[arrow] (sample) -- (estimation);
\draw[arrow] (estimation) -- (firstorder);
\draw[arrow] (firstorder) -- (efficiency);

\end{tikzpicture}

\vspace{0.3cm}
\footnotesize \textbf{Target-first organization}

\end{minipage}

\caption{
Two ways of organizing a first graduate course in statistical inference.
A classical model-first sequence begins with probability models and develops
estimation, testing, and large-sample theory within that structure.
The target-first sequence begins with the scientific question and target parameter,
then proceeds through model choice, observed data, estimation, sampling behavior,
inference, and efficiency.
}
\label{fig:sidebyside_clean}
\end{figure}
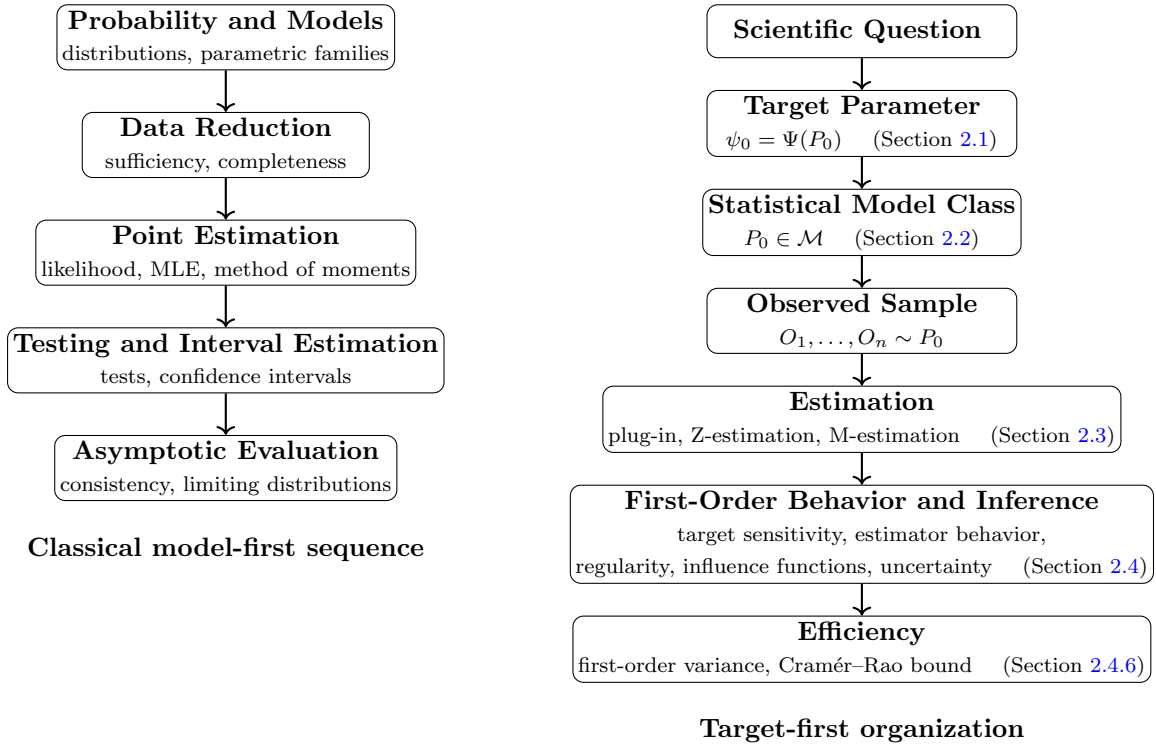

The proposed organization is summarized in \cref{tab:proposal_inference1}, while \cref{fig:sidebyside_clean} contrasts a classical model-first sequence with the target-first workflow. More advanced semiparametric machinery is deferred to a subsequent course, where the same framework can be extended to flexible machine-learning estimators and other modern inferential problems.

To make this progression concrete, \cref{sec:worked-example} follows a simple example to show how familiar classical results emerge naturally from the target-first organization.

\subsection{From Scientific Question to Target Parameter}
\label{sec:target}

The first objective is to translate a scientific question into a precise target of inference. When inference is organized around a model-first progression, the quantity being learned can remain implicit. In a target-first course, students first ask what quantity would answer the scientific question if the relevant population distribution were known.

Formally, let \(O\) denote the observed data for one unit, let \(\mathcal P\) denote a broad collection of distributions for \(O\), and let \(P_0\in\mathcal P\) denote the true data-generating distribution. A target parameter is a functional of this distribution:
\begin{align*}
\Psi : \mathcal{P} \rightarrow \mathbb{R}
\end{align*}
The true target value is
\begin{align*}
\psi_0 = \Psi(P_0)
\end{align*}
This notation separates three objects that students often conflate: the unknown distribution \(P_0\), the target value \(\psi_0\), and the estimator \(\hat\psi\). If \(P_0\) were known, there would be no statistical problem; \(\Psi(P_0)\) could be computed directly. The need for inference arises because \(P_0\) is unknown.

Defining the target parameter requires close collaboration with subject-matter experts and making explicit the outcome of interest, its scale and time point, the population of interest, and, when relevant, the treatment, exposure, or policy comparison, as well as whether the target is causal or associational.

For example, suppose a clinician is interested in how three-year survival varies with a baseline biomarker \(X\) among patients with newly diagnosed cancer. The population of interest is patients with newly diagnosed cancer, the outcome is survival, the time point is three years, and the scale is the probability of survival. Let \(Y^3=\mathbb{I}(T>3)\) denote the indicator of survival beyond three years and let \(P_0\) denote the unknown joint distribution of \(O=(X,Y^3)\) in this population. A natural associational target is
\begin{align*}
m_0(x)
&=
\E_{P_0}\left(Y^3\mid X=x\right)
=
P_{P_0}(T>3\mid X=x).
\end{align*}
Here, \(m_0(x)\) is the target feature of \(P_0\) that answers the scientific question.

Different scientific questions lead to different targets. For example, interest in overall three-year survival would  lead to
\begin{align*}
\psi_0&=\E_{P_0}\left(Y^3\right) = P_{P_0}(T>3)
\end{align*}
while changing the outcome, time point, population, scale, or scientific comparison would define a different target. If the scientific question were instead causal, for example the effect of a new treatment on three-year survival, a corresponding causal estimand might be the risk difference
\begin{align*}
\psi_0^{\mathrm{causal}}&=\E\left(Y^3_1\right)-\E\left(Y^3_0\right)
\end{align*}
where \(Y^3_a\) denotes the counterfactual three-year survival outcome under treatment level \(a\). Unlike the associational targets above, this counterfactual quantity is not generally identified from the observed-data distribution \(P_0\) without additional causal assumptions. Such assumptions would be needed to express the causal estimand as a functional of \(P_0\), but are not developed here (see \cite{pearl2010introduction}).
At this stage, the target has only been defined. The next step is to consider what structure, if any, to impose on the underlying distribution \(P_0\) for studying our target.

\begin{pedagogicalgoal}
\textbf{Pedagogical goal:}
The pedagogical goal is to understand and identify the target of inference, while making explicit the distinction between the unknown data-generating distribution \(P_0\), the target \(\psi_0=\Psi(P_0)\), and an estimator of that target.
\end{pedagogicalgoal}

\subsection{Statistical Model Classes}
\label{sec:models}

Once students have learned to define the target parameter, they can next consider what structure, if any, to impose on the unknown data-generating distribution \(P_0\). Recall that the target \(\psi_0=\Psi(P_0)\) is the feature of \(P_0\) of interest. A statistical model \(\mathcal M\) can then be introduced as the collection of distributions considered possible for the observed data, with \(P_0\in\mathcal M\).

For example, a \textit{parametric model} restricts \(P_0\) to a finite-dimensional family,
\begin{align*}
\mathcal M_{\mathrm{par}} = \left\{ P_\theta:\theta\in\Theta\subseteq\mathbb R^d \right\}
\end{align*}
so functionals of interest can be expressed as functions of a finite-dimensional parameter \(\theta\).

Structure may also be introduced through \textit{moment restrictions}. For a vector-valued function \(g(O,\theta)\), one may consider
\begin{align*}
\mathcal M_{\mathrm{mr}} = \left\{ P\in\mathcal P: \E_P\{g(O,\theta)\}=0 \text{ for some } \theta\in\Theta
\right\}
\end{align*}
Unlike a parametric model, moment restrictions constrain selected features of the distribution rather than the full distribution. This approach provides the basis for method-of-moments and generalized method-of-moments estimation \citep{Hansen1982,NeweyMcFadden1994}.

For the marginal survival target introduced above, specific examples are as follows. One may specify
\(
\psi_0=\E_{P_0}(Y^3),
\)
and impose a parametric model 
\begin{equation}\label{eq:paraY3}
\mathcal M_{\mathrm{par}}^{(Y^3)} = \left\{ \mathrm{Bernoulli}(p):p\in[0,1] \right\},
\qquad
\Psi(P_p)=p
\end{equation}

The same target can also be characterized using the moment-restriction formulation by taking
\(g(O,\psi)=Y^3-\psi\) and \(\theta=\psi\). Then
\[
\mathcal M_{\mathrm{mr}}^{(Y^3)} =
\left\{ P\in\mathcal P: \E_P(Y^3-\psi)=0 \text{ for some }\psi\in[0,1] \right\},
\qquad
\Psi(P)=\psi=\E_P(Y^3)
\]

The same perspective extends to semiparametric and nonparametric models, where less structure is imposed on \(P_0\). These model classes are not developed here, but they illustrate why distinguishing the target from the model is useful: the same target can be studied under increasingly flexible assumptions.

\begin{pedagogicalgoal}
\textbf{Pedagogical goal:}
The pedagogical goal is to present the statistical model as a choice about how much structure to impose on \(P_0\) after the target has been defined. Parametric models and moment restrictions provide two different ways of introducing such structure.
\end{pedagogicalgoal}

\subsection{Estimation of the Target}
\label{sec:estimation}

Students can next learn how a population-level target is connected to observed data through estimation. The target parameter and statistical model remain population-level objects: \(P_0\) is unknown, \(\psi_0=\Psi(P_0)\) specifies the feature of \(P_0\) we wish to learn, and \(P_0\in\mathcal M\) represents the modeling assumptions imposed on \(P_0\). Estimation begins with an observed sample,
\begin{align*}
O_1,\ldots,O_n \overset{\mathrm{iid}}{\sim} P_0
\end{align*}
and constructs an estimator \(\hat\psi=\hat\psi(O_1,\ldots,O_n)\) to approximate \(\psi_0\).
Several broad estimation strategies fit naturally within this framework. A natural approach is the plug-in strategy, which  $\hat\psi=\Psi(\hat P)$, which simply estimates relevant parts of the distribution $P$, then plugs in those estimates to a formula for the target parameter as if we had used the truth. Mathematically, it is described as $ \hat\psi=\Psi(\hat P)$. In the example of \cref{eq:paraY3}, this means estimating $p$ directly using proportions. 

A Z-estimation approach can also be used. Define the empirical measure by $ P_n f=\frac{1}{n}\sum_{i=1}^n f(O_i) $. Then, a Z-estimator \(\hat\theta\) solves $P_n g(O,\hat\theta)=0$ where \(g(O,\theta)\) is a moment or estimating function. An example of this is given by using logistic regression to model \cref{eq:paraY3} in which the estimating function is a minimizer/maximizer of the gradient function. 
Method of moments and estimating equations fit naturally within Z-estimation, with the related M-estimation defining estimators through
optimization of an empirical criterion
\citep{VanDerVaart1998,NeweyMcFadden1994,Godambe1991}, also falling under the class of Z-estimation.

\begin{pedagogicalgoal}
\textbf{Pedagogical goal:}
The pedagogical goal is to present plug-in and Z-estimation as different routes for estimation, yet working toward the same goal of estimating a target.
\end{pedagogicalgoal}

\subsection{Sampling Behavior, Statistical Inference, and Efficiency}
\label{sec:sampling}
Once an estimator \(\hat\psi\) has been constructed, the next objective is to teach students how the target and estimator behave under small changes in the data-generating distribution, how these behaviors are linked, and how this connection leads to uncertainty quantification and efficiency. We begin with the population-level sensitivity of the target, study the first-order sampling behavior of the estimator, and finally connect the two through regularity and influence functions.

\subsubsection{Population-Level Sensitivity: Local Perturbations and Scores}
\label{sec:pop-level-sens}

Students can first learn how the target changes under small, smooth perturbations of the data-generating distribution. To describe such perturbations, consider a smooth one-dimensional path given by
\begin{align*}
\{P_\epsilon:\epsilon\in(-\delta,\delta)\}\subseteq\mathcal M
\end{align*}
subject to $P_{\epsilon=0}=P_0$. 
This represents a smooth local variation of the distribution, indexed by the scalar \(\epsilon\), with \(P_{\epsilon=0}=P_0\). Along this path, the target becomes a function of \(\epsilon\), namely \(\psi(P_\epsilon)\), whose derivative at zero describes the sensitivity to this perturbation. 

To rigorously quantify how distributions change along this submodel, the \textit{score} is introduced by
$$ S(O) = \left. \frac{\partial}{\partial \epsilon} \log(p_\epsilon(O)) \right|_{\epsilon=0}$$

An example of such a submodel is given by 
$$ p_\epsilon(o) = p_0(o) (1+\epsilon S(o))$$
where for parametric models, the score is the usual derivative of the log-likelihood.

Finally, to describe how this translates the target parameter, the pathwise derivative of $\psi$ at $P_0$ is introduced along the submodels $P_\epsilon$, \citep{VanDerVaart1998,Bickel1993,Tsiatis2006}
$$ \dot\Psi_{P_0}(S) = \left. \frac{\partial}{\partial\epsilon}\Psi(P_\epsilon) \right|_{\epsilon=0} $$

\begin{pedagogicalgoal}
\textbf{Pedagogical goal:}
The pedagogical goal is to introduce the score as describing a local direction of perturbation of \(P_0\), and the pathwise derivative as describing the corresponding first-order change in the target, rather than presenting the score as an object tied specifically to maximum likelihood.
\end{pedagogicalgoal}

\subsubsection{First-Order Representation of the Estimation Error}
\label{sec:first-order}

Students can next learn how to study the sampling behavior of an estimator through a common first-order representation of its estimation error. 
Letting $P_n f = \frac{1}{n}\sum_{i=1}^n f(O_i)$ be the empirical measure, consider the following expansion
\begin{align*}
\hat\psi-\psi_0 &= (P_n-P_0)\phi + R_n
\end{align*}
where $\phi$ is mean 0, $\E_{P_0}\{\phi(O)\}=0$, and \(R_n\) collects the non-first order error.
The function \(\phi\) plays a crucial role in determining the estimator's first-order sampling behavior, referred to as the \textit{influence function} of the estimator \citep{Hampel1986,VanDerVaart1998,Bickel1993}. 

This representation provides a common framework for familiar large-sample properties of a general target parameter. If the leading term and remainder vanish in probability, the estimator is consistent. 
If $R_n=o_p(n^{-1/2})$,  the estimator is \textit{asymptotically linear}  \citep{VanDerVaart1998,Bickel1993}, and
\begin{align*}
(P_n-P_0)\phi &= \frac{1}{n}\sum_{i=1}^n\phi(O_i)
\end{align*}
which determines the first-order rate and limiting behavior by the central limit theorem.

\begin{pedagogicalgoal}
\textbf{Pedagogical goal:}
The pedagogical goal is to provide a common first-order representation of the estimation error, from which students can study various statistical properties.
\end{pedagogicalgoal}

\subsubsection{Linking Population-Level Sensitivity and Estimator-Level Behavior:
Regularity and Influence Functions}
\label{sec:regularity-if}

Students can next learn how the first-order sensitivity of the target and the first-order sampling behavior of an estimator are connected. The former is described through the pathwise derivative, while the latter is described through the estimator's asymptotic linear expansion.
Specifically, under \textit{regularity}, these two first-order behaviors are linked by
\[
\dot\Psi_{P_0}(S) = E_{P_0}\!\left[\phi_{P_0}(O)S(O)\right]
\]
where \(\phi_{P_0}\) is the influence function in the expansion of a regular asymptotically linear estimator. Regularity requires this identity to hold for every score \(S\) of a regular submodel through \(P_0\) \citep{Bickel1993,VanDerVaart1998,Tsiatis2006}.

Loosely speaking, regularity means that the estimator's limiting distribution, centered at the perturbed target, remains stable under local \(n^{-1/2}\)-scale perturbations of \(P_0\). This ensures that the first-order sampling behavior is compatible with the corresponding first-order change in the target.

This relationship gives students a common interpretation of the central objects:
\begin{align*}
S
&:\ \text{local direction of perturbation of }P_0\\
\dot\Psi_{P_0}(S)
&:\ \text{pathwise derivative, or first-order change in the target}\\
\phi
&:\ \text{influence function describing the estimator's first-order sampling behavior}
\end{align*}


\begin{pedagogicalgoal}
\textbf{Pedagogical goal:}
The pedagogical goal is to introduce the influence function as the object that links a local direction of perturbation of \(P_0\), represented by the score \(S\), to first-order change in the target and the first-order sampling behavior of a regular estimator via regularity.
\end{pedagogicalgoal}

\subsubsection{From Influence Functions to Statistical Inference}
\label{sec:stat-inf}

Students can next learn how the estimator's first-order representation leads directly to statistical inference.
Under the asymptotic linearity conditions above, the central limit theorem \citep{VanDerVaart1998} gives
\begin{align*}
\sqrt{n}(\hat\psi-\psi_0) &\overset{d}{\longrightarrow} N\left(0,\Var_{P_0}\{\phi(O)\}\right)
\end{align*}
so first order, $\Var(\hat\psi) \approx \frac{1}{n}\Var_{P_0}\{\phi(O)\} $. 
Thus, the influence function directly determines the asymptotic variance of the estimator.
Writing
\begin{align*}
\hat\sigma^2 &= \frac{1}{n-1}\sum_{i=1}^n \left\{\hat\phi(O_i)-(\frac{1}{n}\sum_{i=1}^n\hat\phi(O_i))\right\}^2
\end{align*}
a large-sample confidence interval is then given by a Wald-type construction \citep{LehmannRomano2005}. Let the Wald statistic be given by 
\[
Z = \frac{\hat\psi-\psi^\ast} {\widehat{\mathrm{se}}(\hat\psi)}, \quad \widehat{\mathrm{se}}(\hat\psi)=\sqrt{\frac{\hat {\sigma}^2}{n}}
\]
Then, students can form an asymptotically valid confidence interval by
\[
\hat\psi \pm 1.96 \widehat{\mathrm{se}}(\hat\psi)
\]

which also connects to a hypothesis test and p-value for
\[
    H_0:\psi_0=\psi^\ast
\]

\begin{pedagogicalgoal}
\textbf{Pedagogical goal:}
The pedagogical goal is to show how standard errors, confidence intervals, and hypothesis tests follow from the estimator's first-order sampling behavior within a common inferential framework.
\end{pedagogicalgoal}

\subsubsection{Resampling and the Jackknife}

Students can next learn how resampling provides a computational way to study the sampling behavior of an estimator. Once a target and estimator have been specified, the bootstrap approximates the estimator's sampling distribution by repeatedly applying the same estimation procedure to resampled data \citep{Efron1982,EfronTibshirani1993}. Typically, bootstrap is seen as a computational route to standard errors, confidence intervals, and other features of sampling behavior when analytic approximations are difficult, but the connection to estimation theory can be made explicit and clear using this approach.

The jackknife offers a related connection to the influence-function perspective \citep{boos2013essential}. For sufficiently smooth estimators, the change induced by deleting one observation provides a finite-sample approximation to that observation's first-order influence on the estimator. Jackknife pseudovalues are viewed as empirical analogues of influence-function contributions, linking classical resampling ideas to the first-order representation developed above.

\begin{pedagogicalgoal}
\textbf{Pedagogical goal:}
The pedagogical goal is to present the bootstrap and jackknife from the basis of the target first approach, providing an alternative to studying the sampling behavior of a specified estimator. In particular, the jackknife can be used to reinforce the connection between finite-sample perturbations and influence functions.
\end{pedagogicalgoal}

\subsubsection{Efficiency}
\label{sec:efficiency}

Once students understand that the influence function determines first-order variance, $\Var_{P_0}\{\phi(O)\}$, 
efficiency can be introduced as the question of how small this variance can be among regular estimators of the same target \citep{LehmannCasella1998,VanDerVaart1998,Bickel1993,Tsiatis2006}.

Parametric models provide a concrete setting in which to develop this idea. For a model \(P_\theta\), let the score vector be given by
\begin{align*}
S_\theta(O) &= \frac{\partial}{\partial\theta} \log p_\theta(O)
\end{align*}
and the Fisher information by
\begin{align*}
I(\theta) &= \E_\theta\left\{ S_\theta(O)S_\theta(O)^\top \right\}
\end{align*}

The one-dimensional score corresponding to a local direction through the parametric model is obtained from this score vector. For a smooth scalar target \(\psi(\theta)\), the \textit{efficient influence function} in the parametric model is
\begin{align*}
\phi_{\mathrm{eff}}(O) &= \nabla\psi(\theta_0)^\top I(\theta_0)^{-1} S_{\theta_0}(O)
\end{align*}
with variance
\begin{align*}
\Var_{P_0}\{\phi_{\mathrm{eff}}(O)\} &= \nabla\psi(\theta_0)^\top I(\theta_0)^{-1} \nabla\psi(\theta_0)
\end{align*}
recovering the Cramér--Rao lower bound \citep{LehmannCasella1998,CasellaBerger2002}. One can leverage much of the classical pedagogy to present efficiency in this view, and proceed to discuss equivalent ideas when moving past parametric models. 

\begin{pedagogicalgoal}
\textbf{Pedagogical goal:}
The pedagogical goal is to frame efficiency as a comparison of regular estimators through their first-order variance, and to present the likelihood score, Fisher information, and Cramér--Rao lower bound as the parametric realization of these more general ideas.
\end{pedagogicalgoal}

\subsection{Extensions of the Target-First Framework}
\label{sec:extensions}
We have presented an outline for a target first framework, going from the scientific question to target, the model class assumptions, the estimator choice, and an overview of understanding the sampling behavior, inference, and efficiency. Moving forward, we discuss two useful and natural extensions after developing this foundation: multiple testing, where the inferential target becomes multivariate, and regression, where the distinction between the target and the model becomes especially important.

\subsubsection{Multiple Targets and Multiple Testing}

Once students have learned to formulate inference around a single target parameter, the same framework can be extended to a collection of targets,
\[
\boldsymbol{\psi}_0=\left\{\Psi_1(P_0),\ldots,\Psi_K(P_0)\right\}
\]
The inferential problem then includes not only estimating each target and quantifying its uncertainty, but also controlling errors across the collection. Multiple-testing procedures can therefore be introduced as an extension of the same target-first sequence, with family-wise error rate and false discovery rate providing different ways to formalize error control across several inferential targets \citep{Holm1979,BenjaminiHochberg1995}.

\begin{pedagogicalgoal}
\textbf{Pedagogical goal:}
The pedagogical goal is to extend the target-first framework from a single inferential target to a collection of targets and to motivate multiple-testing procedures through the need to control uncertainty across that collection.
\end{pedagogicalgoal}

\subsubsection{Regression and Working Models}

Once students have learned to distinguish the target parameter from the statistical model and estimator, the same framework can be used to revisit regression from a target-first perspective. Regression is often introduced by selecting a model based on the outcome type, such as linear regression for continuous outcomes or logistic regression for binary outcomes, and then interpreting the resulting model coefficients. However, these coefficients may not correspond to the original scientific question, and their usual interpretations may depend on correct model specification.

A target-first approach instead begins by asking which population quantity is of scientific interest and then considers how regression can be used to summarize or estimate that quantity. This provides two complementary ways of introducing regression within the same inferential framework.

First, regression can be used to define a lower-dimensional summary of a potentially complex population relationship. For example, consider the population conditional mean
\[
    m_0(x)=\E_{P_0}(Y\mid X=x)
\]
without assuming a particular functional form. Rather than postulating that this relationship is linear, we may define the coefficients of its best linear approximation:
\[
    \beta^\ast = \arg\min_{\beta} \E_{P_0} \left[\{m_0(X)-\beta_0-\beta_1X\}^2\right]
\]
The corresponding working model is
\[
    m_\beta(x)=\beta_0+\beta_1x
\]
Here, $\beta^\ast$ is a well-defined population quantity, provided the projection is unique, even when the true conditional mean is nonlinear. The coefficients describe a particular summary of the population relationship rather than parameters of an assumed true linear model \citep{White1982,Buja2019}. Their scientific relevance must still be justified by the
research question.

Second, regression can be used as an estimation tool without its coefficients being the target of inference. For example, suppose the scientific question concerns the marginal mean
\[
    \psi_0 =  \E_{P_0}(Y) = \E_{P_0}\!\left\{\E_{P_0}(Y\mid X)\right\}
\]
A regression model for the conditional mean
\(\E_{P_0}(Y\mid X)\) can be used to help estimate \(\psi_0\),
but the target remains the marginal mean rather than any individual regression coefficient.

These examples illustrate how familiar regression methods can be incorporated into the target-first framework without requiring the regression model to determine the scientific target. Regression can provide a useful population summary or serve as a tool for estimating a separately defined target, depending on the scientific question.

\begin{pedagogicalgoal}
\textbf{Pedagogical goal:}
The pedagogical goal is to extend the target-first framework to regression by teaching students to distinguish the scientific target from the role of the regression model. Regression coefficients may be defined as population-level summaries without assuming correct model specification, while regression models may also be used to estimate targets that are not themselves regression coefficients.
\end{pedagogicalgoal}

\subsection{A Bayesian Route to Inference}
\label{sec:bayesian}
Once students have learned to separate the scientific question, target parameter, statistical model, estimator, and uncertainty quantification, the same framework provides a natural way to introduce basic Bayesian inference. The scientific question, target parameter, statistical model, and observed data remain the same; Bayesian inference introduces a prior distribution on the unknown parameters of the model and updates this distribution using the observed data.

For example, under a parametric model \(P_\theta\), a prior \(\pi(\theta)\) and likelihood \(p_\theta(O_1,\ldots,O_n)\) lead to the posterior \citep{Gelman2013}
\[
    \pi(\theta\mid O_1,\ldots,O_n)
    \propto p_\theta(O_1,\ldots,O_n)\pi(\theta)
\]
If the target is \(\psi=\Psi(P_\theta)\), the posterior for \(\theta\) induces a posterior distribution for the target, which can be summarized using posterior means, credible intervals, or posterior probabilities.

This presentation emphasizes the common structure of the inferential problem rather than framing Bayesian and frequentist methods as competing approaches. Both begin with the same scientific question and target, but they differ in how uncertainty about that target is constructed and interpreted.

\begin{pedagogicalgoal}
\textbf{Pedagogical goal:}
The pedagogical goal is to use the common target-first structure to introduce Bayesian inference as an alternative route to uncertainty quantification.
\end{pedagogicalgoal}

\section{A Target-First Course Design}

The framework above describes organizing principles rather than a syllabus.
In this section, we illustrate how it could be implemented in a 14-week first graduate course in statistical inference.
\Cref{tab:proposal_inference1} summarizes the main components of the proposed organization and their pedagogical roles, while \cref{tab:weeks_inference1} gives one possible week-by-week implementation.
The course develops the target-first sequence through estimation, first-order sampling behavior, inference, resampling, and efficiency, and then uses the same framework to revisit multiple testing, regression and working models, and basic Bayesian inference. The final component provides connections to a subsequent course in modern statistical inference.
Concepts such as pathwise differentiability, regularity, and influence functions can be introduced at a conceptual and computational level, while more technical semiparametric theory and efficiency calculations are deferred to subsequent coursework.

\begin{table}[h!]
\centering
\caption{A target-first organization for a first graduate course in statistical inference.}

\footnotesize
\setlength{\tabcolsep}{3pt}
\renewcommand{\arraystretch}{1.04}

\begin{tabularx}{\textwidth}{
    @{}
    >{\raggedright\arraybackslash}p{0.035\textwidth}
    >{\raggedright\arraybackslash}p{0.245\textwidth}
    >{\raggedright\arraybackslash}X
    @{}
}
\toprule
& \textbf{Topic} & \textbf{Pedagogical goal} \\
\midrule

I. & Scientific questions and target parameters
& Foreground the target of inference and distinguish the unknown distribution \(P_0\), the target \(\psi_0=\Psi(P_0)\), and an estimator of that target \\

II. & Statistical model classes
& Present the statistical model as a choice about how much structure to impose on \(P_0\) after the target has been defined \\

III. & Estimation
& Present plug-in, and M-/Z-estimation as alternative routes from a specified target and model to an estimator within a common inferential framework \\

IV. & Population-level sensitivity
& Introduce the score as a local direction of perturbation of \(P_0\) and the pathwise derivative as the corresponding first-order change in the target \\

V. & First-order sampling behavior
& Organize consistency, rates of convergence, and asymptotic distributions around the structure of the estimation error \\

VI. & Regularity and influence functions
& Connect local perturbations of \(P_0\), first-order changes in the target, and the first-order sampling behavior of regular estimators \\

VII. & Statistical inference
& Show how standard errors, confidence intervals, and hypothesis tests follow from the estimator's first-order sampling behavior \\

VIII. & Bootstrap and jackknife
& Present resampling as a computational approach to studying sampling behavior, while using the jackknife to connect finite-sample perturbations with influence functions \\

IX. & Efficiency
& Frame efficiency as a comparison of regular estimators through first-order variance, with Fisher information and the Cramér--Rao bound as the parametric realization of these ideas \\

X. & Multiple targets and multiple testing
& Extend the target-first framework from a single target to a collection of targets and motivate multiple-testing procedures through the need to control inferential errors across that collection \\

XI. & Regression and working models
& Apply the target-first framework to regression, distinguishing regression coefficients as population-level summaries from regression models used to estimate separately defined targets, without requiring correct specification of the working model \\

XII. & A Bayesian route to inference
& Use the common target-first structure to introduce Bayesian inference as an alternative route to uncertainty quantification, while making explicit the role of the prior and the interpretation of uncertainty \\

XIII. & Bridges to modern statistical inference
& Show how the same organizing ideas provide a foundation for flexible machine-learning estimators, policy learning, anytime-valid inference, and modern Bayesian methods \\

\bottomrule
\end{tabularx}

\label{tab:proposal_inference1}
\end{table}

\begin{table}[h!]
\centering
\caption{One possible 14-week implementation of the target-first framework.}

\footnotesize
\setlength{\tabcolsep}{3pt}
\renewcommand{\arraystretch}{1.03}

\begin{tabularx}{\textwidth}{
    @{}
    >{\raggedright\arraybackslash}p{0.055\textwidth}
    >{\raggedright\arraybackslash}p{0.265\textwidth}
    >{\raggedright\arraybackslash}X
    @{}
}
\toprule
\textbf{Week} & \textbf{Topic} & \textbf{Representative material} \\
\midrule

1 & Scientific questions, targets, and models
& Observed data, populations, target parameters, parametric models, moment restrictions \\

2 & General approaches to estimation
& Plug-in estimation, Z-estimation, M-estimation \\

3 & Classical estimators in the common framework
& Maximum likelihood, method of moments, generalized method of moments, estimating equations \\

4 & First-order estimation error
& Error decompositions, LLN, CLT, delta method, consistency, convergence rates \\

5 & Local perturbations and target sensitivity
& Parametric submodels, scores, pathwise derivatives \\

6 & Regularity and influence functions
& Asymptotic linearity, remainder terms, regular estimators, influence functions \\

7 & Statistical inference
& First-order variance, standard errors, confidence intervals, Wald tests \\

8 & Bootstrap
& Bootstrap distributions, standard errors, confidence intervals, relation to sampling distributions \\

9 & Jackknife
& Leave-one-out perturbations, pseudovalues, variance estimation, connection to influence functions \\

10 & Efficiency
& Efficient influence functions, likelihood scores, Fisher information, Cramér--Rao bounds \\

11 & Multiple targets and multiple testing
& Collections of target parameters and hypotheses, family-wise error rate, false discovery rate \\

12 & Regression and working models
& Conditional-mean targets, parametric regression models, regression coefficients, working models, low-dimensional summaries \\

13 & A Bayesian route to inference
& Priors, likelihoods, posterior distributions, posterior summaries, credible intervals \\

14 & Bridges to modern statistical inference
& Flexible ML- and AI-based estimation, policy learning, anytime-valid inference, modern Bayesian methods \\

\bottomrule
\end{tabularx}

\label{tab:weeks_inference1}
\end{table}
\section{Bridging the Target-First Framework to a Second Course in Inference}

The target-first framework provides a natural foundation for a subsequent course in modern statistical inference, where the same ideas of target specification, model choice, first-order approximation, and uncertainty quantification can be extended to more complex settings. We discuss the presentation and possible topics.

\paragraph{Flexible estimation with machine learning and AI.}
Flexible estimation provides a natural setting in which to revisit valid inference and introduce influence-function-based corrections, orthogonality, and cross-fitting \citep{hines2022demystifying,kennedy2024semiparametric,williams2026riesz,Chernozhukov2018}.

\paragraph{Policy learning and decision making.}
Replacing a fixed target with the value of a treatment rule or policy creates a bridge to individualized treatment rules, dynamic treatment regimes, contextual bandits, reinforcement learning, and regret-based decision making \citep{Dudik2011,Kallus2018,AtheyWager2021}.

\paragraph{Anytime-valid inference.}
Sequential monitoring and adaptive stopping provide a natural extension of fixed-sample uncertainty quantification to confidence sequences and anytime-valid inference \citep{ramdas2023game}.

\paragraph{Modern Bayesian inference.}
The same target-first perspective can also connect to modern Bayesian methods, including amortized approaches such as Prior-Data Fitted Networks \citep{Muller2022PFN}.

\paragraph{Higher-Order Influence Functions} 
For nonlinear functionals, higher-order influence functions and classical U- and V-statistics provide natural extensions of the first-order framework. This can be casted in the framework of above, providing exposure to non-standard functional targets.

\section{Discussion}

The goal of the proposed framework is not to replace classical statistical inference, but to reorganize its central ideas around a common inferential workflow. Beginning with the scientific question and target parameter makes the distinction among the target, statistical model, estimator, and uncertainty explicit. Estimation, first-order sampling behavior, influence functions, resampling, and efficiency can then be introduced as successive components of the same inferential problem.

An important advantage of this organization is that the same framework provides a natural bridge to modern statistical inference. The questions developed here, including what the target is, what structure is imposed, how it is estimated, and how uncertainty is characterized, continue to apply in subsequent courses when estimation uses machine learning or artificial intelligence, when the target becomes a treatment policy or decision rule, when inference is sequential, or when uncertainty is represented through a Bayesian posterior.

This target-first perspective is closely related in spirit to the targeted-learning literature, which organizes estimation and inference around a prespecified target parameter and the statistical model in which it is defined \citep{VanDerLaanRose2011}. Our emphasis here is pedagogical: we use this organizing principle to connect classical and modern inferential methods rather than to advocate for a particular estimation procedure.

This reorganization necessarily changes the relative emphasis placed on some topics traditionally covered in mathematical statistics courses. Classical results on data reduction and finite-sample optimality, including sufficiency and minimal sufficiency, completeness, Rao--Blackwell and Lehmann--Scheff\'e results, and Neyman--Pearson or uniformly most powerful testing theory, may receive less attention. These remain valuable ideas, and programs wishing to retain exposure to them could reserve a lecture for selected classical finite-sample results while preserving the target-first organization of the course.

Finally, this is a curricular proposal rather than an empirical evaluation of learning outcomes, and the effects of this organization on student learning remain to be studied.

\paragraph{Acknowledgements}
We acknowledge Ivan Diaz for the helpful discussions.


\bibliographystyle{apalike}

\bibliography{bib}

@article{kennedy2024semiparametric,
  title={Semiparametric doubly robust targeted double machine learning: a review},
  author={Kennedy, Edward H},
  journal={Handbook of statistical methods for precision medicine},
  pages={207--236},
  year={2024},
  publisher={Chapman and Hall/CRC}
}

@article{hines2022demystifying,
  title={Demystifying statistical learning based on efficient influence functions},
  author={Hines, Oliver and Dukes, Oliver and Diaz-Ordaz, Karla and Vansteelandt, Stijn},
  journal={The American Statistician},
  volume={76},
  number={3},
  pages={292--304},
  year={2022},
  publisher={Taylor \& Francis}
}

@article{williams2026riesz,
  title={Riesz representers for the rest of us},
  author={Williams, Nicholas T and Hines, Oliver J and Rudolph, Kara E},
  journal={Epidemiology},
  pages={10--1097},
  year={2026},
  publisher={LWW}
}

@inproceedings{Dudik2011,
  title     = {Doubly Robust Policy Evaluation and Learning},
  author    = {Dud{\'i}k, Miroslav and Langford, John and Li, Lihong},
  booktitle = {Proceedings of the 28th International Conference on Machine Learning},
  pages     = {1097--1104},
  year      = {2011}
}

@book{Bickel1993,
  title     = {Efficient and Adaptive Estimation for Semiparametric Models},
  author    = {Bickel, Peter J. and Klaassen, Chris A. J. and Ritov, Ya'acov and Wellner, Jon A.},
  year      = {1993},
  publisher = {Johns Hopkins University Press}
}

@book{CasellaBerger2002,
  title     = {Statistical Inference},
  author    = {Casella, George and Berger, Roger L.},
  edition   = {2},
  year      = {2002},
  publisher = {Duxbury}
}

@article{Chernozhukov2018,
  title   = {Double/Debiased Machine Learning for Treatment and Structural Parameters},
  author  = {Chernozhukov, Victor and Chetverikov, Denis and Demirer, Mert and Duflo, Esther and Hansen, Christian and Newey, Whitney and Robins, James},
  journal = {The Econometrics Journal},
  volume  = {21},
  number  = {1},
  pages   = {C1--C68},
  year    = {2018}
}

@book{Tsiatis2006,
  title     = {Semiparametric Theory and Missing Data},
  author    = {Tsiatis, Anastasios A.},
  year      = {2006},
  publisher = {Springer}
}

@book{VanDerLaanRose2011,
  title     = {Targeted Learning: Causal Inference for Observational and Experimental Data},
  author    = {van der Laan, Mark J. and Rose, Sherri},
  year      = {2011},
  publisher = {Springer}
}

@book{VanDerVaart1998,
  title     = {Asymptotic Statistics},
  author    = {van der Vaart, A. W.},
  year      = {1998},
  publisher = {Cambridge University Press}
}

@inproceedings{Kallus2018,
  title     = {Balanced Policy Evaluation and Learning},
  author    = {Kallus, Nathan},
  booktitle = {Advances in Neural Information Processing Systems},
  volume    = {31},
  year      = {2018}
}

@article{AtheyWager2021,
  title   = {Policy Learning with Observational Data},
  author  = {Athey, Susan and Wager, Stefan},
  journal = {Econometrica},
  volume  = {89},
  number  = {1},
  pages   = {133--161},
  year    = {2021},
  doi     = {10.3982/ECTA15732}
}

@article{ramdas2023game,
  title     = {Game-Theoretic Statistics and Safe Anytime-Valid Inference},
  author    = {Ramdas, Aaditya and Gr{\"u}nwald, Peter and Vovk, Vladimir and Shafer, Glenn},
  journal   = {Statistical Science},
  volume    = {38},
  number    = {4},
  pages     = {576--601},
  year      = {2023},
  publisher = {Institute of Mathematical Statistics}
}

@inproceedings{Muller2022PFN,
  title     = {Transformers Can Do Bayesian Inference},
  author    = {M{\"u}ller, Samuel and Hollmann, Noah and Pineda Arango, Sebastian and Grabocka, Josif and Hutter, Frank},
  booktitle = {International Conference on Learning Representations},
  year      = {2022}
}

@article{pearl2010introduction,
  title={An introduction to causal inference},
  author={Pearl, Judea},
  journal={The international journal of biostatistics},
  volume={6},
  number={2},
  pages={7},
  year={2010}
}

@article{Hansen1982,
  title   = {Large Sample Properties of Generalized Method of Moments Estimators},
  author  = {Hansen, Lars Peter},
  journal = {Econometrica},
  volume  = {50},
  number  = {4},
  pages   = {1029--1054},
  year    = {1982}
}

@incollection{NeweyMcFadden1994,
  title     = {Large Sample Estimation and Hypothesis Testing},
  author    = {Newey, Whitney K. and McFadden, Daniel},
  booktitle = {Handbook of Econometrics},
  editor    = {Engle, Robert F. and McFadden, Daniel L.},
  volume    = {4},
  pages     = {2111--2245},
  year      = {1994},
  publisher = {Elsevier}
}

@book{Godambe1991,
  title     = {Estimating Functions},
  editor    = {Godambe, V. P.},
  year      = {1991},
  publisher = {Clarendon Press},
  address   = {Oxford}
}

@book{Hampel1986,
  title     = {Robust Statistics: The Approach Based on Influence Functions},
  author    = {Hampel, Frank R. and Ronchetti, Elvezio M. and Rousseeuw, Peter J. and Stahel, Werner A.},
  year      = {1986},
  publisher = {John Wiley \& Sons}
}

@book{Efron1982,
  title     = {The Jackknife, the Bootstrap and Other Resampling Plans},
  author    = {Efron, Bradley},
  year      = {1982},
  publisher = {Society for Industrial and Applied Mathematics}
}

@book{EfronTibshirani1993,
  title     = {An Introduction to the Bootstrap},
  author    = {Efron, Bradley and Tibshirani, Robert J.},
  year      = {1993},
  publisher = {Chapman \& Hall}
}

@book{LehmannCasella1998,
  title     = {Theory of Point Estimation},
  author    = {Lehmann, E. L. and Casella, George},
  edition   = {2},
  year      = {1998},
  publisher = {Springer}
}

@book{LehmannRomano2005,
  title     = {Testing Statistical Hypotheses},
  author    = {Lehmann, E. L. and Romano, Joseph P.},
  edition   = {3},
  year      = {2005},
  publisher = {Springer}
}

@article{White1982,
  title   = {Maximum Likelihood Estimation of Misspecified Models},
  author  = {White, Halbert},
  journal = {Econometrica},
  volume  = {50},
  number  = {1},
  pages   = {1--25},
  year    = {1982}
}

@article{Buja2019,
  title   = {Models as Approximations I: Consequences Illustrated with Linear Regression},
  author  = {Buja, Andreas and Berk, Richard and Brown, Lawrence and George, Edward and Pitkin, Emil and Traskin, Mikhail and Zhang, Kai and Zhao, Linda},
  journal = {Statistical Science},
  volume  = {34},
  number  = {4},
  pages   = {523--544},
  year    = {2019}
}

@article{Holm1979,
  title   = {A Simple Sequentially Rejective Multiple Test Procedure},
  author  = {Holm, Sture},
  journal = {Scandinavian Journal of Statistics},
  volume  = {6},
  number  = {2},
  pages   = {65--70},
  year    = {1979}
}

@article{BenjaminiHochberg1995,
  title   = {Controlling the False Discovery Rate: A Practical and Powerful Approach to Multiple Testing},
  author  = {Benjamini, Yoav and Hochberg, Yosef},
  journal = {Journal of the Royal Statistical Society: Series B},
  volume  = {57},
  number  = {1},
  pages   = {289--300},
  year    = {1995}
}

@book{boos2013essential,
  title={Essential statistical inference: theory and methods},
  author={Boos, Dennis D and Stefanski, Leonard A},
  volume={591},
  year={2013},
  publisher={Springer}
}

@book{Gelman2013,
  title     = {Bayesian Data Analysis},
  author    = {Gelman, Andrew and Carlin, John B. and Stern, Hal S. and Dunson, David B. and Vehtari, Aki and Rubin, Donald B.},
  edition   = {3},
  year      = {2013},
  publisher = {Chapman \& Hall/CRC}
}


\clearpage
\appendix

\section{A Worked Example of the Target-First Framework}
\label{sec:worked-example}

This appendix illustrates the proposed organization using a deliberately simple example. 
Suppose the scientific question concerns three-year survival among patients with newly diagnosed cancer. 
Let
\[
Y^3=\mathbb I(T>3)
\]
indicate survival beyond three years. 
To keep the calculations transparent, we initially take \(O=Y^3\); a baseline biomarker \(X\) is reintroduced in the extensions below. 
The simplicity of the example allows the same problem to be followed from target specification through estimation, first-order behavior, inference, resampling, and efficiency.

\subsection*{Scientific Question and Target}

The scientific question is: what proportion of patients in the population survive beyond three years? 
Following \cref{sec:target}, the corresponding target is
\[
\psi_0 = \Psi(P_0) = \E_{P_0}(Y^3) = P_{P_0}(T>3)
\]
where \(P_0\) denotes the unknown population distribution of \(Y^3\).

At this stage, the target has been defined without specifying either a statistical model or an estimator.

\noindent\textit{Pedagogical goal illustrated (\cref{sec:target}):}
separate the unknown distribution \(P_0\), the target \(\psi_0=\Psi(P_0)\), and the estimator that will later be used to approximate it.

\subsection*{Statistical Model}

Because \(Y^3\) is binary, its marginal distribution can be represented as
\[
Y^3\sim\mathrm{Bernoulli}(p), \qquad p\in[0,1]
\]
so that
\[
\mathcal M = \{\mathrm{Bernoulli}(p):p\in[0,1]\}, \qquad \psi(p)=p
\]
In this example the Bernoulli family exhausts the possible distributions on \(\{0,1\}\), so the parametric representation imposes no additional restriction beyond the binary support.

\noindent\textit{Pedagogical goal illustrated (\cref{sec:models}):}
the target is specified first, while the statistical model describes the structure imposed on the unknown distribution.

\subsection*{Estimation}

Suppose
\[
Y_1^3,\ldots,Y_n^3
\overset{\mathrm{iid}}{\sim}P_0
\]
The natural estimator is the sample proportion
\[
\hat\psi=P_nY^3 = \frac{1}{n}\sum_{i=1}^nY_i^3
\]

This single estimator illustrates the three estimation strategies introduced in \cref{sec:estimation}. 
First, it is a plug-in estimator:
\[
\hat\psi
=
\Psi(P_n).
\]
Second, it is a Z-estimator because it solves
\[
    P_n(Y^3-\hat\psi)=0
\]
Third, under the Bernoulli model it is the maximum likelihood estimator,
\[
\hat\psi = \arg\max_{p\in[0,1]} P_n \left[ Y^3\log p+(1-Y^3)\log(1-p) \right]
\]
and hence an M-estimator with negative log-likelihood as its criterion.

Thus, plug-in estimation, Z-estimation, and M-estimation provide different representations of the same familiar estimator.

\noindent\textit{Pedagogical goal illustrated (\cref{sec:estimation}):}
place familiar estimation methods within a common inferential workflow and show how plug-in, Z-estimation, and M-estimation can provide different representations of the same estimator.

\subsection*{Population-Level Sensitivity}

Following \cref{sec:sampling}, consider a local path through the true Bernoulli distribution,
\[
p_\epsilon=p_0+\epsilon h
\]
for sufficiently small \(\epsilon\). 
The corresponding score is
\[
S_h(Y^3) = h\, \frac{Y^3-p_0}{p_0(1-p_0)}.
\]
Because the target is \(\psi(p)=p\),
\[
\dot\Psi_{P_0}(S_h) = \left. \frac{\partial}{\partial\epsilon} \psi(p_\epsilon) \right|_{\epsilon=0} = h
\]

The score therefore describes the local direction in which \(P_0\) is perturbed, while the pathwise derivative describes the corresponding first-order change in the survival probability.

\noindent\textit{Pedagogical goal illustrated (\cref{sec:sampling}):}
connect a local perturbation of the data-generating distribution to the resulting first-order change in the target.

\subsection*{First-Order Sampling Behavior}

For the sample proportion,
\[
\hat\psi-\psi_0 = P_nY^3-P_0Y^3 = (P_n-P_0)(Y^3-\psi_0)
\]
Thus the general first-order representation becomes
\[
\hat\psi-\psi_0 = (P_n-P_0)\phi,
\qquad
\phi(Y^3)=Y^3-\psi_0
\]
with
\[
R_n=0
\]

In this simple example, the first-order representation is exact, and
\(\phi(Y^3)=Y^3-\psi_0\) is the function governing the estimator's
first-order sampling behavior. The next step is to verify that this same
function is compatible with the local behavior of the target.
Consistency follows from the law of large numbers, while the central limit theorem determines the root-\(n\) limiting behavior.

\noindent\textit{Pedagogical goal illustrated (\cref{sec:sampling}):}
organize consistency, rates, and limiting distributions around the structure of the estimation error.

\subsection*{Regularity and the Influence Function}

The function identified from the first-order representation can now be
connected to the population-level perturbation. For the path indexed by \(h\),
\[
\E_{P_0} \left[ \phi(Y^3)S_h(Y^3) \right] = \E_{P_0} \left[ (Y^3-p_0) h\frac{Y^3-p_0}{p_0(1-p_0)} \right] = h
\]
Hence
\[
\dot\Psi_{P_0}(S_h) = \E_{P_0}\{\phi(Y^3)S_h(Y^3)\}
\]

Thus the same \(\phi(Y^3)=Y^3-\psi_0\) that describes the estimator's
first-order sampling behavior also represents the pathwise derivative through
the influence function relationship. The example therefore makes explicit the
connection among the score, the first-order change in the target, and the
first-order sampling behavior of the estimator described in \cref{sec:sampling}.

\noindent\textit{Pedagogical goal illustrated (\cref{sec:sampling}):}
show concretely how the influence function links population-level sensitivity to the first-order sampling behavior of a regular estimator.

\subsection*{Statistical Inference}

Since
\[
\Var_{P_0}\{\phi(Y^3)\} = \psi_0(1-\psi_0)
\]
the central limit theorem gives
\[
\sqrt n(\hat\psi-\psi_0) \overset{d}{\longrightarrow} N\left\{0,\psi_0(1-\psi_0)\right\}
\]
The corresponding estimated standard error is
\[
\widehat{\mathrm{se}}(\hat\psi) = \sqrt{ \frac{\hat\psi(1-\hat\psi)}{n} }
\]
A large-sample confidence interval is therefore
\[
\hat\psi \pm 1.96\,\widehat{\mathrm{se}}(\hat\psi)
\]
and for \(H_0:\psi_0=\psi^\ast\), the corresponding Wald statistic is
\[
Z = \frac{\hat\psi-\psi^\ast} {\widehat{\mathrm{se}}(\hat\psi)}
\]

Thus standard errors, confidence intervals, and hypothesis tests arise directly from the first-order representation.

\noindent\textit{Pedagogical goal illustrated (\cref{sec:sampling}):}
derive familiar inferential procedures from the estimator's first-order sampling behavior.

\subsection*{Bootstrap and Jackknife}

The same estimator also illustrates the role of resampling. 
A nonparametric bootstrap sample
\[
Y_1^{3\ast},\ldots,Y_n^{3\ast}
\]
is drawn from the empirical distribution \(P_n\), and the estimator is recomputed as
\[
\hat\psi^\ast = \frac{1}{n}\sum_{i=1}^nY_i^{3\ast}
\]
The conditional distribution of \(\hat\psi^\ast-\hat\psi\) provides a computational approximation to the sampling distribution of \(\hat\psi-\psi_0\)

For the jackknife, deleting observation \(i\) gives
\[
\hat\psi_{(-i)} = \frac{n\hat\psi-Y_i^3}{n-1}
\]
The corresponding jackknife pseudovalue is
\[
J_i = n\hat\psi-(n-1)\hat\psi_{(-i)} = Y_i^3
\]
so that
\[
J_i-\hat\psi = Y_i^3-\hat\psi
\]
The centered pseudovalues therefore coincide with the empirical version of the influence-function contribution.

\noindent\textit{Pedagogical goal illustrated (\cref{sec:sampling}):}
present bootstrap and jackknife methods as tools for studying the sampling behavior of the same estimator, while making the connection between jackknife perturbations and influence functions explicit.

\subsection*{Efficiency}

For the Bernoulli model, the likelihood score is
\[
S_p(Y^3) = \frac{Y^3-p}{p(1-p)}
\]
and the Fisher information is
\[
I(p) = \frac{1}{p(1-p)}
\]
Because \(\psi(p)=p\),
\[
\phi_{\mathrm{eff}}(Y^3) = I(p_0)^{-1}S_{p_0}(Y^3) = Y^3-p_0
\]
Thus
\[
\Var_{P_0}\{\phi_{\mathrm{eff}}(Y^3)\} = p_0(1-p_0)
\]
and the corresponding variance bound for an estimator based on \(n\) observations is
\[
\frac{p_0(1-p_0)}{n}
\]

The influence function of the sample proportion is therefore the efficient influence function, and the estimator attains the Cramér--Rao bound in this model.

\noindent\textit{Pedagogical goal illustrated (\cref{sec:sampling}):}
connect first-order variance, likelihood scores, Fisher information, and the Cramér--Rao bound as different parts of the same efficiency calculation.

\end{document}